\documentstyle[12pt] {article}
\newcommand{\beq}{\begin{equation}}
\newcommand{\eeq}{\end{equation}}
\newcommand{\beqa}{\begin{eqnarray}}
\newcommand{\eeqa}{\end{eqnarray}}
\newcommand{\ba}{\begin{array}}
\newcommand{\ea}{\end{array}}

\begin{document}

\begin{center}
{\large \bf Instabilities, Point Attractors and Limit Cycles \\
in a Inflationary Universe} \\
\end{center}

\vskip 1. truecm

\begin{center}
{\bf Luca Salasnich}\footnote{Electronic address:
salasnich@padova.infn.it}
\vskip 0.5 truecm
Dipartimento di Fisica "G. Galilei" dell'Universit\`a di Padova, \\
Via Marzolo 8, I 35131 Padova, Italy \\
and\\
Departamento de Fisica Atomica, Molecular y Nuclear \\
Facultad de Ciencias Fisicas, Universidad "Complutense" de Madrid, \\
Ciudad Universitaria, E 28040 Madrid, Spain\\
\end{center}

\vskip 1. truecm

\begin{center}
{\bf Abstract}
\end{center}

\vskip 0.5 truecm
\par
We study the stability of a scalar inflaton field and
analyze its point attractors in the phase space.
We show that the value of the inflaton field in the vacuum
is a bifurcation parameter and prove the possible existence
of a limit cycle by using analytical and numerical arguments.

\begin{center}
To be published in Modern Physics Letters A
\end{center}

\newpage

\par
The phenomenon of chaos$^{1}$, which has become very popular, rejuvenated
interest in nonlinear dynamics, and the chaotic behaviour of classical
field theories is currently subject of intensive research$^{2,3,4,5}$.
In this respect it is of great interest to investigate the existence
of some precursor phenomena of chaotic motion. In particular
nonlinear effects should lead to bifurcations which qualitatively change
the system properties$^{6,7,8}$.
\par
In this paper we study the stability of a scalar inflaton
field$^{9,10,11}$ and analyze its bifurcation properties
in the framework of the dynamical system theory$^{6,7,8}$.
The model we study is very schematic, so it can be seen
as a toy model for classical nonlinear dynamics,
with the attractive feature that it emerges form particle
physics and cosmology.
\par
It is generally believed that the universe, at a very early stage after
the big bang, exhibited a short period of exponential expansion, the
so--called inflationary phase$^{9,10,11}$.
In fact the assumption of an inflationary universe
solves three major cosmological problems: the flatness problem, the
homogeneity problem, and the formation of structure problem.
\par
The Friedmann--Robertson--Walker metric$^{12}$ of a homogeneous and isotropic
expanding universe is given by:
\beq
ds^2=dt^2-a^2(t)\big[ {dr^2\over 1 - k r^2}+r^2(d\theta^2 +
\sin^2{\theta} d\varphi^2 )\big],
\eeq
where $k=1,-1$, or $0$ for a closed, open, or flat universe, and $a(t)$
is the scale factor of the universe.
\par
The evolution of the scale factor $a(t)$ is given by
the Einstein equations:
$$
{\ddot a}=-{4\pi \over 3}G(\rho + 3 p)a ,
$$
\beq
\big({{\dot a} \over a}\big)^2+{k\over a^2}=
{8\pi \over 3}G \rho ,
\eeq
where $\rho$ is the energy density of matter in the universe,
and $p$ its pressure. The gravitational constant $G=M_p^{-2}$
(with $\hbar =c=1$),
where $M_p =1.2 \cdot 10^{19}$ GeV is the Plank mass, and
$H={\dot a}/a$ is the Hubble "constant", which in general is a function
of time.
\par
All inflationary models postulate the existence of a scalar field
$\phi$, the so--called inflation field, with Lagrangian$^{13}$:
\beq
L={1\over 2}\partial_{\mu}\phi \partial^{\mu}\phi -V(\phi )
\eeq
where the potential $V(\phi )$ depends on the type of inflation model
considered. Here we choose a real field but also
complex scalar can be used$^{11}$. The scalar field,
if minimally coupled to gravity, satisfies the equation$^{9,10,11}$:
\beq
\Box \phi = {\ddot \phi} + 3 \big({\dot a\over a}\big)
{\dot \phi} - {1\over a^2} \nabla^2 \phi
= -{\partial V \over \partial \phi},
\eeq
where $\Box$ is the covariant d'Alembertian operator.
We suppose that in the universe there is only the inflaton field,
so the Hubble "constant" $H$ is related to the energy density of
the field by:
\beq
H^2 +{k\over a^2}=
\big({{\dot a} \over a}\big)^2+{k\over a^2}=
{8 \pi G \over 3} \big[
{{\dot \phi}^2 \over 2}+{(\nabla \phi )^2\over 2}+ V(\phi )
\big].
\eeq
\par
Immediately after the onset of inflation, the cosmological scale factor
grows exponentially$^{9,10,11}$. Thus the term $\nabla^2 \phi /a^2$ is
generally believed to be negligible and, if the inflaton field is
sufficiently uniform (i.e. ${\dot \phi}^2$, $(\nabla \phi )^2 << V(\phi )$),
we end up with a classical nonlinear scalar field theory in one dimension:
\beq
{\ddot \phi } + 3 H(\phi ) {\dot \phi} +
{\partial V \over \partial \phi}=0,
\eeq
where the Hubble "constant" $H$ is an explicit function of $\phi$:
\beq
H^2 = {8 \pi G \over 3} V(\phi ).
\eeq
Most inflation scenarios are just formulated in this classical setting.
In fact the quantization of the inflation scenario can be regarded to
be still an open problem$^{14}$.
\par
The classical inflaton dynamics can be studied in the framework
of the dynamical system theory$^{6,7,8}$.
The second order equation of motion can be
written as a system of two first order differential equations:
\beq
{\dot {\vec x}}={\vec g}({\vec x})
\eeq
where ${\vec x}=(\phi , \chi )$ is a point in the two dimensional
phase space and ${\vec g}=(g_1,g_2)$ is given by:
\beq
g_1(\phi ,\chi )=\chi ,
\quad
g_2(\phi ,\chi )= - 3 H(\phi ) \chi
- {\partial V(\phi )\over \partial \phi} .
\eeq
The system is non--conservative because the function
\beq
div({\vec g})=
{\partial g_1 \over \partial \phi }+{\partial g_2 \over \partial \chi}
=- 3 H(\phi )
\eeq
is not identically zero. The fixed points of the system are those for which
$g_1(\phi ,\chi )=0$ and $g_2(\phi ,\chi )=0$, i.e:
\beq
\chi =0,
\quad
{\partial V(\phi )\over \partial \phi}=0 .
\eeq
\par
The deviation $\delta {\vec x}(t)={\hat {\vec x}}(t)-{\vec x} (t)$
from the two initially neighboring trajectories
${\vec x}$ and ${\hat{\vec x}}$ in the phase space
satisfies the linearized equations of motion:
\beq
{d \over dt}\delta {\vec x}(t)= \Gamma (t) \delta {\vec x}
\eeq
where $\Gamma (t)$ is the stability matrix:
\beq
\Gamma (t)= {\left(\matrix {0 & 1 \cr
- {\partial^2 V\over \partial \phi^2} - 3\chi {\partial H\over \partial \phi}
& - 3H(\phi ) \cr } \right) }.
\eeq
At least if an eigenvalue of $\Gamma (t)$ is real
the separation of the trajectories grows exponentially
and the motion is unstable. Imaginary eigenvalues correspond to stable motion.
In the limit of time that goes to infinity the eigenvalues of the stability
matrix are the Lyapunov exponent$^{1}$. It is well known that for
two--dimensional dynamical system the Lyapunov exponent can not be positive
and so the system is not chaotic, i.e. there is not global instability.
\par
However, we can be assured that the universe is crowded with many
interacting fields of which the inflaton is but one$^{11}$.
The nonlinear nature of these interactions can result in a complex
chaotic evolution of the universe and the local instability
of the inflaton field is a precursor phenomenon of chaotic motion.
Following the Toda criterion$^{15}$, we assume that the time dependence
can be eliminated, i.e. $\Gamma (\phi (t),\chi (t))=\Gamma (\phi ,\chi )$.
\par
The eigenvalues of the stability matrix are given by:
\beq
\sigma_{1,2}=-{3\over 2}H(\phi )\pm {1\over 2}\sqrt{9 H^2(\phi )
-4 {\partial^2 V\over \partial \phi^2}
-12 \chi {\partial H \over \partial \phi} }.
\eeq
The pair of eigenvalues become real and there
is exponential separation of neighboring trajectories, i.e.
unstable motion, if:
\beq
{\partial^2 V \over \partial \phi^2} +
3\chi {\partial H\over \partial \phi}< 0.
\eeq
Particularly when $\chi =0$, e.g. the fixed points, we obtain
local instability when:
\beq
{\partial^2 V \over \partial \phi^2} < 0,
\eeq
i.e. for negative curvature of the potential energy. The fixed points
are stable if they are point of local minimum of $V(\phi )$ and unstable
if are points of local maximum.
\par
As stressed at the beginning the potential $V(\phi )$ depends on the
type of inflation model considered, and it is usually some
kind of double--well potential. We choose a symmetric double--well
potential:
\beq
V(\phi )={\lambda \over 4}(\phi^2 -v^2)^2 ,
\eeq
where $\pm v$ are the values of the inflaton field in the vacuum,
i.e. the points of minimal energy of the system.
\par
We observe that the inflaton field value in the vacuum $v$ is a
bifurcation parameter. Bifurcation is used to indicate a qualitative
change in the features of the system under the variation of one or more
parameters on which the system depends$^{6,7,8}$.
First of all we consider the case $v=0$, i.e. $V(\phi )=(\lambda /4) \phi^4$.
In this situation there is only one fixed point $(\phi^*=0,\chi^*=0)$
which is a stable one being:
\beq
{\partial^2 V\over \partial \phi^2}=3\lambda \phi^2 \geq 0.
\eeq
The fixed point $(\phi^*=0,\chi^*=0)$ is a point attractor, as shown
in Fig. 1 where we use a fourth order Runge--Kutta method$^{16}$
to compute the classical trajectories in the phase space.
\par
Instead for $v\neq 0$ there are three fixed points:
\beq
(\phi^* =0,\chi^* =0), \quad (\phi^* = v,\chi^* =0) ,
\quad (\phi^* =-v,\chi^* =0),
\eeq
and the condition for the instability becomes:
\beq
-{v\over \sqrt{3}}< \phi < {v\over \sqrt{3}}.
\eeq
Obviously $(\phi^* =0,\chi^* =0)$ is an unstable fixed point, and
in particular a saddle point because the stability matrix has
real and opposite eigenvalues.
On the other hand $(\phi^* =\pm v, \chi^* =0)$ are stable
fixed points.
\par
Form eq. (7) we find four possible functions for the Hubble "constant":
\beq
H(\phi )= \pm \gamma |\phi^2-v^2| ,
\eeq
but also:
\beq
H(\phi )= \pm \gamma (\phi^2-v^2) ,
\eeq
where $\gamma=\sqrt{2\pi G\lambda / 3}$ is the dissipation parameter.
The choice of the Hubble function is crucial for the dynamical
evolution of the system.
\par
In certain non--conservative systems we could find closed trajectories or limit
cycles toward which the neighboring trajectories spiral on both sides.
It is sometime possible to know that no limit cycle exist and
the Bendixson criterion$^{17}$, which establishes a condition for the
non--existence of closed trajectories, is useful in some cases.
Bendixson criterion is as follows: if $div({\vec g})$ is not zero and
does not change its sign within a domain $D$ of the phase space,
no closed trajectories can exist in that domain.
\par
In our case we have $div({\vec g})=- 3 H(\phi )$,
and so the presence of periodic orbit is related to the sign of $H(\phi )$.
If $H(\phi )=\gamma |\phi^2-v^2|$ we do not find periodic orbits
and the inflaton field goes to one of its two stable fixed points,
which are points attractors (see Fig. 2). The vacuum is degenerate but
if we choose an initial condition around the saddle point
there is a dynamical symmetry breaking$^{13}$ towards the positive $v$
or negative $-v$ value of the inflaton field in the vacuum.
This symmetry breaking is unstable because neighbour
initial conditions can go in different point attractors.
\par
Instead, if we choose $H(\phi )=\gamma (\phi^2-v^2)$ the numerical
calculations of Fig. 3 show that exists a limit cycle,
the two stable fixed points are not point attractors,
and the inflaton field oscillates forever.
Obviously more large is $v$ more large is the limit cycle.
\par
In conclusion, we have studied the stability of a scalar
inflaton field. With a symmetric double--well self energy
the value of the inflaton field in the vacuum is a bifurcation parameter
which changes dramatically the phase space structure.
The main point is that for some functional solutions of the Hubble "constant"
the system goes to a limit cycle, i.e. to a periodic orbit.
\par
The inflaton field is not chaotic but its
local instability can give rise to a complex chaotic evolution of the
universe due to its nonlinear interactions with other fields.
In the future it will be very interesting to study these effects
which can perhaps lead to some observable implications
like a fractal pattern in the spectrum of density fluctuations.

\vskip 0.5 truecm

\begin{center}
*****
\end{center}
\par
The author is greatly indebted to Prof. S. Graffi, Prof. V. R. Manfredi
and Dr. A. Riotto for many enlightening discussions.
The author acknowledges Prof. J. M. G. Gomez for his hospitality
at the Department of Atomic, Molecular and Nuclear Physics
of "Complutense" University, and the "Ing. Aldo Gini" Foundation
for a partial support.

\newpage

\begin{center}
{\bf References}
\end{center}

\vskip 0.5 truecm

\noindent

1. A. J. Lichtenberg and M. A. Lieberman, {\it Regular and Stochastic
Motion} (Springer--Verlag, Berlin, 1983);
M. C. Gutzwiller, {\it Chaos in Classical and Quantum Mechanics}
(Springer--Verlag, Berlin, 1991).

2. G. K. Savvidy, Phys. Lett. B {\bf 130}, 303 (1983);
G. K. Savvidy, Nucl. Phys. B {\bf 246}, 302 (1984).

3. T. Kawabe and S. Ohta, Phys. Rev. D {\bf 44}, 1274 (1991);
T. Kawabe and S. Ohta, Phys. Lett. B {\bf 334}, 127 (1994);
T. Kawabe, Phys. Lett. B {\bf 343}, 254 (1995).

4. M. S. Sriram, C. Mukku, S. Lakshmibala and B. A. Bambah, Phys. Rev. D
{\bf 49}, 4246 (1994).

5. S. Graffi, V. R. Manfredi and L. Salasnich, Mod. Phys. Lett. B {\bf 7},
747 (1995); L. Salasnich, Phys. Rev. D {\bf 52}, 6189 (1995).

6. C. Hayashi, {\it Nonlinear Oscillations in Physical Systems}
(Princeton University Press, Princeton, 1985).

7. J. Awrejecewicz, {\it Bifurcations and Chaos in Coupled Oscillators}
(World Scientific, Singapore, 1991).

8. A. H. Nayfeh and B. Balachandran, {\it Applied Nonlinear Dynamics}
(J. Wiley, New York, 1995).

9. A. H. Guth, Phys. Rev. D {\bf 23} B, 347 (1981).

10. A. D. Linde, Phys. Lett. B {\bf 108}, 389 (1982);
A. D. Linde, Phys. Lett. B {\bf 129}, 177 (1983).

11. A. D. Linde, {\it Particle Physics and Inflationary Cosmology}
(Harwood Academic Publishers, London, 1988).

12. S. Weinberg, {\it Gravitation and Cosmology} (Wiley, New York, 1972).

13. C. Itzykson and J. B. Zuber, {\it Quantum Field Theory}
(McGraw--Hill, New York, 1985).

14. R. H. Brandenberger, in SUSSP Proceedings {\it Physics of the Early
Universe}, Eds. J.A. Peacock, A. F. Heavens and A. T. Daves (Institute of
Physics Publishing, Bristol, 1990).

15. M. Toda, Phys. Lett. A {\bf 48}, 335 (1974);
G. Benettin, R. Brambilla and L. Galgani, Physica A {\bf 87}, 381 (1977).

16. Subroutine D02BAF, The NAG Fortran Library, Mark 14, Oxford: NAG
Ltd. and USA: NAG Inc. (1990).

17. I. Bendixson, Acta Math. {\bf 24}, 1 (1901).

\newpage

\parindent=0.pt

\section*{Figure Captions}
\vspace{0.6 cm}

{\bf Figure 1}: Inflaton field {\it vs} time (up),
and its phase space trajectory (down), with $\gamma =1/2$,
$\lambda =1$ and $v=0$.
Initial conditions: $\phi =0$ and $\chi ={\dot \phi}=1/2$.

{\bf Figure 2}: Inflaton field {\it vs} time (up),
and its phase space trajectory (down),
for $H(\phi )=\gamma |\phi^2-v^2|$ with $\gamma =1/2$, $\lambda =1$
and $v=1$. Initial conditions: $\phi =0$ and $\chi ={\dot \phi}=1/2$.

{\bf Figure 3}: Inflaton field {\it vs} time (up),
and its phase space trajectory (down),
for $H(\phi )=\gamma (\phi^2-v^2)$ with $\gamma =1/2$, $\lambda =1$
and $v=1$. Initial conditions: $\phi =0$ and $\chi ={\dot \phi}=1/2$.

\end{document}